# Ultrahigh reversible hydrogen storage in K and Ca decorated 4-6-8 biphenylene sheet

*Vikram Mahamiya[a*], Alok Shukla[a], Brahmananda Chakraborty[b,c*]*

[a]Indian Institute of Technology Bombay, Mumbai 400076, India

[b]High pressure and Synchrotron Radiation Physics Division, Bhabha Atomic Research Centre, Bombay, Mumbai, India-40085

[c]Homi Bhabha National Institute, Mumbai, India-400094

email: vikram.physics@iitb.ac.in ; shukla@phy.iitb.ac.in ; brahma@barc.gov.in

## Abstract

By applying density functional theory (DFT) and *ab-initio* molecular dynamics (AIMD) simulations, we predict the ultrahigh hydrogen storage capacity of K and Ca decorated single-layer biphenylene sheet (BPS). We have kept various alkali and alkali-earth metals, including Na, Be, Mg, K, Ca, at different sites of BPS and found that K and Ca atoms prefer to bind individually on the BPS instead of forming clusters. It was found that 2×2×1 supercell of biphenylene sheet can adsorb eight K, or eight Ca atoms, and each K or Ca atom can adsorb 5 $H_2$, leading to 11.90 % or 11.63 % of hydrogen uptake, respectively, which is significantly higher than the DOE-US demands of 6.5 %. The average adsorption energy of $H_2$ for K and Ca decorated BPS is -0.24 eV and -0.33 eV, respectively, in the suitable range for reversible $H_2$ storage. Hydrogen molecules get polarized in the vicinity of ionized metal atoms hence get attached to the metal atoms through electrostatic and van der Waals interactions. We have estimated the desorption temperatures of $H_2$ and found that the adsorbed $H_2$ can be utilized for reversible use. We have found that a sufficient energy barrier of 2.52 eV exists for the movement of Ca atoms, calculated using the climbing-image nudged elastic band (CI-NEB) method. This energy barrier can prevent the clustering issue of Ca atoms. The solidity of K and Ca decorated BPS structures were investigated using AIMD simulations.

## 1. Introduction

Hydrogen is considered one of the most suitable green energy sources that can replace fossil fuels in automotive applications[1]. Safe, compact, and efficient hydrogen storage are some of the serious challenges for the scientific community in present times[2,3]. Large bulky pressure tanks are required to store hydrogen in the gaseous form, which involves safety concerns. Transportation is also one of the major issues for hydrogen storage in the gaseous phase. Due to the high liquefaction cost, liquid-phase hydrogen storage is not recommended[4]. The solid-state form of hydrogen storage is suitable if the substrate material can store hydrogen with high gravimetric density and hydrogen storage is reversible. The department of energy, United States (DOE-US)[5,6], has issued few guidelines for practically suitable hydrogen storage materials. The binding energy of $H_2$ should lie in the range of -0.1 eV to -0.7 eV[7], and hydrogen uptake should be more than 6.5 %. However, the average binding energy of the $H_2$ should lie in the range of -0.2 eV/$H_2$ to -0.4 eV/$H_2$ for reversible hydrogen storage[8,9].

Different kinds of substrate materials, for example, metal hydrides and alloys[10–15], metal-organic frameworks[16–18], zeolites[19,20], have been explored for hydrogen storage. However, there are serious issues with these substrate materials, such as high desorption temperature for metal hydrides, lower hydrogen uptake for zeolites, instability at high temperatures, clustering of the metal atoms, etc. Metal doped carbon nanomaterials such as fullerenes[8,21–24], carbon nanotubes[25–34], graphene[35–39], graphyne[40–43], advanced 2d materials[44–47] have also been studied widely for hydrogen storage due to their low molecular mass and high surface area. Pristine carbon nanomaterials are not suitable for hydrogen storage as they bind the hydrogen molecules only by the weak van der Waals forces at ambient conditions[48,49], hence desorption temperature is lower than the room temperature. Although transition metal (TM) decorated carbon nanomaterials can adsorb many hydrogen molecules on a single metal atom through Kubas interactions[50,51], the issue is that the possibilities of metal clustering are significant in TMs decorated carbon nanomaterials due to the large cohesive energy of TMs[8]. The metal-metal clustering can lower the hydrogen uptake to a great extent. Alkali and alkali-earth metals (AM and AEM) have much lower cohesive energy than the TMs, so the chances of the metal clustering are minimal[52–54]. The adsorption energies of the hydrogen molecules attached on AM or AEM doped substrates are generally lesser than the TM doped substrates and suitable for the reversible use of hydrogen[23,55]. In addition to that, high hydrogen uptake can be achieved due to the lower molecular mass of AMs and AEMs compared to TMs. Hydrogen storage capability in Li, Na, and Ca doped $C_{24}$ fullerene was examined by Zhang et al.[23]. They have predicted high hydrogen uptake of up to 12.7 % for their systems. Recently reversible hydrogen storage capabilities of Sc decorated $C_{24}$ fullerene system were investigated by Mahamiya et al.[56]. They have predicted a very high 13.02 % of hydrogen uptake. Hydrogen storage properties of yttrium decorated $C_{24}$ fullerene are also investigated by Mahamiya et al.[57]. They have reported that each yttrium atom can adsorb 6 hydrogen molecules reversibly by Kubas interactions leading to 8.84 wt % of hydrogen. Li et al.[8] found that the Sc and Ti atoms form a cluster when doped on $B_{80}$ surface. They have found 8.2% of hydrogen uptake for Ca doped $B_{80}$ fullerene. Lee et al.[54,58] have found that hydrogen molecules are bonded with Ca decorated carbon nanostructures by s-d hybridization (Kubas interactions), which is absent in Mg decorated carbon nanostructures.

Sahoo et al.[59] have studied hydrogen storage properties of Li and Na decorated $C_{20}$ fullerene. They have found that each Li and Na atom attached to $C_{20}$ fullerene can bind 5 $H_2$ molecules leading up to 13.08 wt % of hydrogen. Beheshti et al.[60] have found 8.37 % of hydrogen uptake for boron-doped Ca decorated graphene structure. Ataca et al.[39] have investigated hydrogen storage capabilities of Ca decorated graphene system. They have found that one Ca atom can bind 5 $H_2$, leading to 8.4 % of hydrogen uptake. Ultrahigh hydrogen storage capacity of 18.6 wt % for Li decorated graphyne was theoretically predicted by Guo et al.[43]. Gangan et al.[42] have investigated hydrogen storage properties of the yttrium doped graphyne system. Borophene and boron substituted substrates are also proven to be high-capacity hydrogen storage materials. Chen et al.[61] have reported up to 9.5 wt % of hydrogen for Ca decorated borophene[61]. Aydin et al.[62] have explored hydrogen storage in Li, Na, and Mg decorated $BC_3$-graphene systems. Eroglu et al.[63] have studied the effect of boron substitution on double carbon vacancy (DCV) graphene. Zhou et al.[64] have reported that the hydrogen storage capacity of graphene increases with boron doping.

Gao et al.[47] have estimated 12.8 wt % of hydrogen uptake for Li doped newly synthesized material holey graphyne recently. Hydrogen adsorption and desorption properties of scandium decorated holey graphyne were recently investigated by Mahamiya et al.[65] by using density functional theory and ab-initio molecular dynamics simulations.

There have been various studies on the production of pure hydrogen as well. Hydrogen production by electrolysis of water proton exchange membrane (PEM) was reported by Grigoriev et al.[66]. Dincer and Zamfirescu[67] have suggested various methods for sustainable hydrogen production, including water splitting methods and extracting hydrogen with other materials than water. Ibrahim Dincer and Canan Acar have reviewed hydrogen production methods from renewable and non-renewable sources for suitable sustainability[67–69]. Ibrahim Dincer has also explore different green methods for hydrogen production[70].

The 4-6-8 membered biphenylene sheets (BPS) were synthesized on a large scale employing East-West expansion of n-phenylenes with different lengths by Schlutter et al.[71] in 2014. The biphenylene sheets have six, four, and eight-membered carbon rings. Hudspeth et al.[72] have studied the electronic properties of the BPS structure and its derivative in one dimension. They have found that the BPS structure is metallic, which remains metallic in the planner strips with zigzag-type edges. However, armchair-edged strips get a bandgap that decreases continuously with the width of the ribbon. Pablo A. Denis[73] has proposed that the bandgap of the metallic BPS structure can be opened and regulated with the doping of halogen (F, Cl) functional groups. Brunetto et al.[74] have shown that a new carbon allotrope biphenylene carbon (BPC) could be formed by selective dehydrogenation of graphene structure. Rahaman et al.[75] have found that by applying uniaxial loading to penta-graphene, which is semiconducting in nature, it can be transformed into metallic BPS. They have also proposed that by heating the BPS structure at a very high temperature (5000 K), it can be transformed into a hexa-graphene structure. Due to their large surface area, these sheets have also been studied for energy storage. The Li-ion storage capacity of biphenylene membrane was explored by Ferguson et al.[76]. They found that biphenylene membrane is suitable for Li-ion battery anode. Recently biphenylene network sheet was experimentally synthesized by an on-surface interpolymer dehydrofluorination (HF-zipping) reaction[77]. Except for the Li doped BPS structure, the hydrogen storage properties of this material have not been studied up to now to the best of our knowledge. Denis et al.[78] have found that Li doped BPS structure can adsorb up to 7.4 wt % of hydrogen, but the average adsorption energy of $H_2$ is -0.20 eV, so the desorption of hydrogen molecules will occur at room temperature itself, rendering the system completely useless for normal operations. In addition to that, they have also not investigated the stability and clustering issues in Li decorated BPS at high temperatures.

We have investigated the reversible $H_2$ adsorption and desorption properties of K and Ca decorated BPS structures by using DFT and AIMD simulations. K and Ca atoms are attached strongly to the BPS due to the charge transfer from metal atoms to BPS. We have presented the density/partial density of states (DOS/PDOS) and spatial charge density difference plots to explain the charge transfer and orbital interactions between metal atoms and BPS. Bader charge analysis[79] calculations are carried out to get the exact amount of charge transfer. Hydrogen molecules are bonded to K and Ca cations through electrostatic interactions along with van der Waals interactions. In addition to that, orbital hybridization between vacant 3d orbitals of Ca atom and σ orbitals of $H_2$ are also responsible for hydrogen binding. The presence of a sufficient energy barrier for the metal atoms and structural solidity at desorption temperature makes our system practically viable for hydrogen storage. These are some of the crucial aspects of this study since diffusion energy barrier calculations and molecular dynamics simulations were not performed in most of the previous reports on hydrogen storage.

## 2. Computational details

We have carried out the geometry optimization calculations using DFT implemented in Vienna Ab Initio Simulation package (VASP)[80–83], along with an exchange-correlation functional employing the generalized gradient approximation (GGA). A 2×2×1 supercell of BPS containing 24 carbon atoms was used for the calculations, and 10 Å of vacuum space is given to avoid the interactions between the two consecutive periodic layers of BPS. A Monkhorst-Pack *k*-grid of 5×5×1 kpoints was taken to sample the Brillouin zone. The convergence limit for the Hellman-Feynman forces and energy is set to be 0.01 eV/Å and $10^{-5}$ eV, respectively. The kinetic energy cut-off for the plane-wave basis expansion is taken to be 500 eV. The DFT-GGA results are corrected using Grimme's DFT-D2[84,85] dispersion corrections, as the DFT-GGA method does not correctly account for the weak van der Waals interactions. The AIMD[86] calculations have been performed to check the stability of metal decorated BPS structures at high desorption temperatures.

## 3. Results and discussions

### 3.1 Interaction and bonding of alkali and alkali-earth metals (AM and AEM) on BPS

We have considered 2×2×1 supercell of BPS structure for the hydrogen storage calculations as presented in **Fig. 1(a).** The unit cell of the BPS structure is presented in **Fig. 1(b).** We have placed various AMs (Na, K) and AEMs (Be, Mg, Ca) at different sites of BPS structure. The metal atoms have been placed at different sites such as T (above the center of the tetragon), H (above the center of the hexagon), and O (above the center of the octagon) as displayed in **Fig. 1(a)** and relaxation calculations were performed**.**

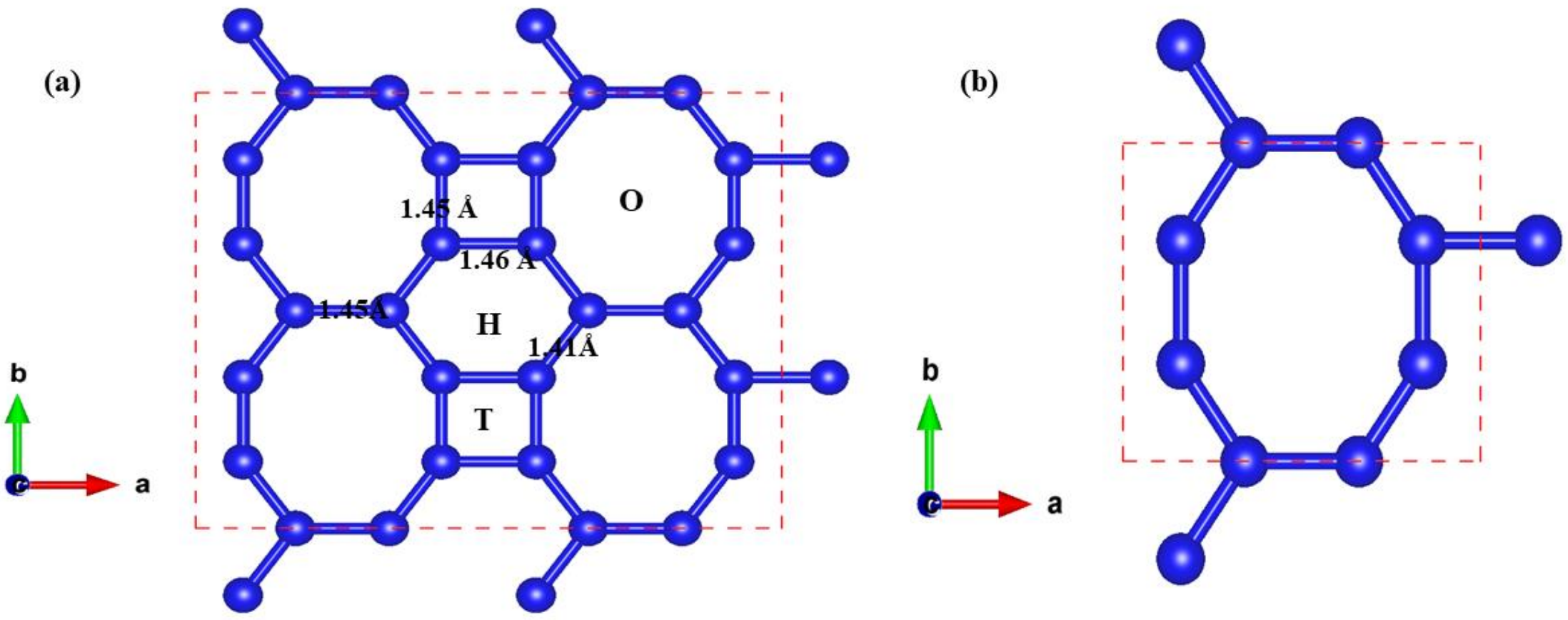

**Fig. 1. The optimized structures (a) 2×2×1 supercell of BPS with 24 carbon atoms. T, H and O represents the center of tetragon, hexagon and octagon of BPS. (b) Unit cell of BPS. Blue color sphere denotes the carbon atoms.**

We have found that the Ca atom placed on T and H sites move slightly during relaxation and come above the common side of the tetragon and hexagon (T-H), while the metal atoms (K and Ca) placed on the O site remain on the top of the octagon of the BPS structure after the relaxation. The metal atoms placed on some other sites come near to these sites (T, H, O) after relaxation. We have found the Na, Be, Mg atoms bind on the BPS structure with very small binding energy (less than 1 eV), and the cohesive energy of these metals is significantly higher than their binding energy on the BPS structure. Hence, these metals will prefer to form clusters instead of binding individually on the BPS. Therefore, we have not considered BPS structures decorated by these metals for the $H_2$ adsorption calculations. The maximum binding energy of the K atom attached to the BPS structure (1.14 eV) is more than the cohesive energy of bulk K (0.93 eV)[87], therefore, the K atom prefers to attach individually to the BPS structure. The maximum binding energy of the Ca atom attached on BPS (1.41 eV) is in between the cohesive energy of cluster Ca (1.30 eV) and bulk Ca (1.84 eV)[88,89]. Hence, we have also carried out the diffusion energy barrier calculation and molecular dynamics simulations to confirm the absence of clustering in Ca decorated BPS structure.

The binding energies of the K and Ca atoms at different sites of the BPS structure are given in **Table 1**.

**Table 1. Binding energy of K and Ca atoms at different sites of BPS structure. T, H, O, and T-H denote center of tetragon, hexagon, octagon, and common side of tetragon and hexagon, respectively.**

| Decorated metal atom | Initial site on BPS | Final site after relaxation | Binding energy of metal atom (eV) |
|---|---|---|---|
| **K** | T | T | -1.06 |
| | H | H | -1.12 |
| | O | O | -1.14 |
| **Ca** | T | T-H | -1.41 |
| | H | T-H | -1.41 |
| | O | O | -1.25 |

We have calculated the binding energy of K and Ca metal atoms attached to BPS by the following equation:

$$\boldsymbol{E_b(M) = E\,(BPS + M) - E(BPS) - E(M)} \qquad (1)$$

Where ***E (BPS + M), E(BPS),*** and ***E (M)*** are the energy of the metal decorated BPS, pristine BPS, and isolated metal atom, respectively. Since T and H sites are the minimum energy sites for the Ca decoration on the BPS structure and the O site is the minimum energy site for K decoration, we have considered these structures for hydrogen storage calculations. The relaxed geometries of K and Ca decorated BPS structures at O, T, and H sites are presented in **Fig. 2.** The relaxed geometry for K decorated BPS structure is presented in **Fig. 2(a)**. K atom was placed at almost 2 Å height above the center of the octagon (O) of the BPS structure, and relaxation calculations were performed. The K-C atom bond length is 2.98 Å after relaxation. **Fig. 2(b & c)** are the relaxed geometric structures of Ca decorated BPS. We have kept the Ca atom at almost 2 Å height above the center of tetragon (T) and hexagon (H) positions, and relaxation was performed.

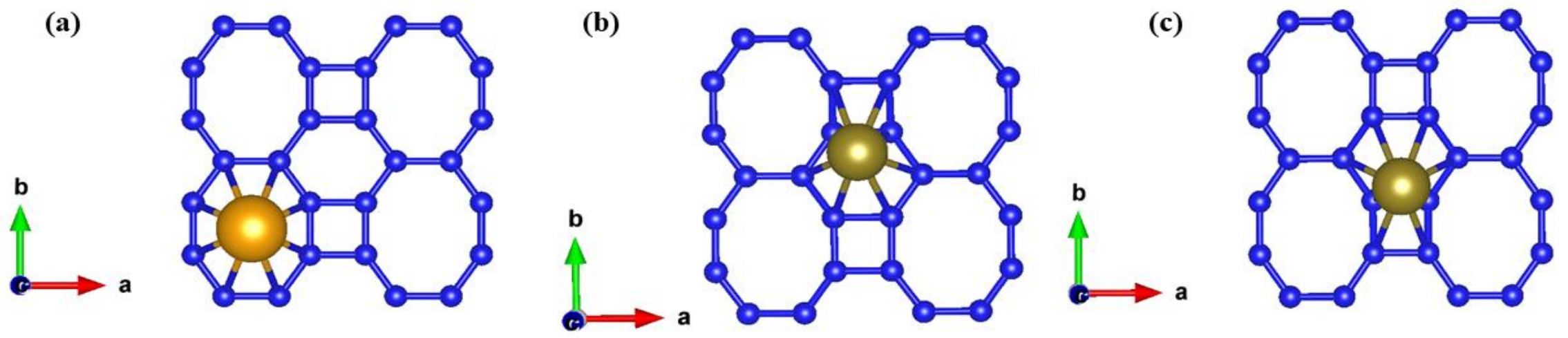

**Fig. 2. The optimized structures of (a) BPS + K, K atom is placed above the center of the octagon (O) before relaxation (b) BPS + Ca, Ca atom is placed above the center of the tetragon (T) before relaxation (c) BPS + Ca, Ca atom is placed above the center of the hexagon (H) before relaxation. Here, blue, orange and golden spheres denote the C, K, and Ca atoms, respectively.**

We have found that the Ca atom placed on T and H sites comes to T-H site (top of the common face of tetragon and hexagon of BPS) after the relaxation. The bond length of Ca to the nearest C atom (Ca-C) of the BPS structure is 2.37 Å after relaxation.

**Density of states (DOS) and partial density of states (PDOS) analysis**

We have presented the total density of states of BPS, K decorated BPS, and Ca decorated BPS in **Fig. 3.** It is clear from **Fig. 3(a)** that the BPS structure is metallic, as reported previously[72,77]. The up and down panels of **Fig. 3(a, b & c)** are symmetric, which denotes that the BPS structure is non-magnetic. The BPS structure remains metallic and non-magnetic after the decoration of K and Ca atoms, however, we can notice that DOS for K and Ca decorated BPS looks different from the DOS of pristine BPS structure. This indicates that there exists some bonding between the metal atoms and the BPS structure. To investigate the orbital bonding and charge transfer phenomena, we have plotted the PDOS of s-orbital of isolated K and Ca atoms and s-orbital of K and Ca atoms when they are attached to the BPS structure, as shown in **Fig. 4.** In **Fig. 4(a),** we can observe some energy states near Fermi level of the PDOS plot of isolated K atom, while these states are missing when K is decorated on BPS structure as shown in **Fig. 4(b).** The depletion of states near the Fermi level is also noticeable in **Fig. 4(d)** compared to **Fig.4(c).** This indicates charge transfer from the s-orbital of K and Ca atoms to the BPS structure when K and Ca atoms are attached to the BPS structure.

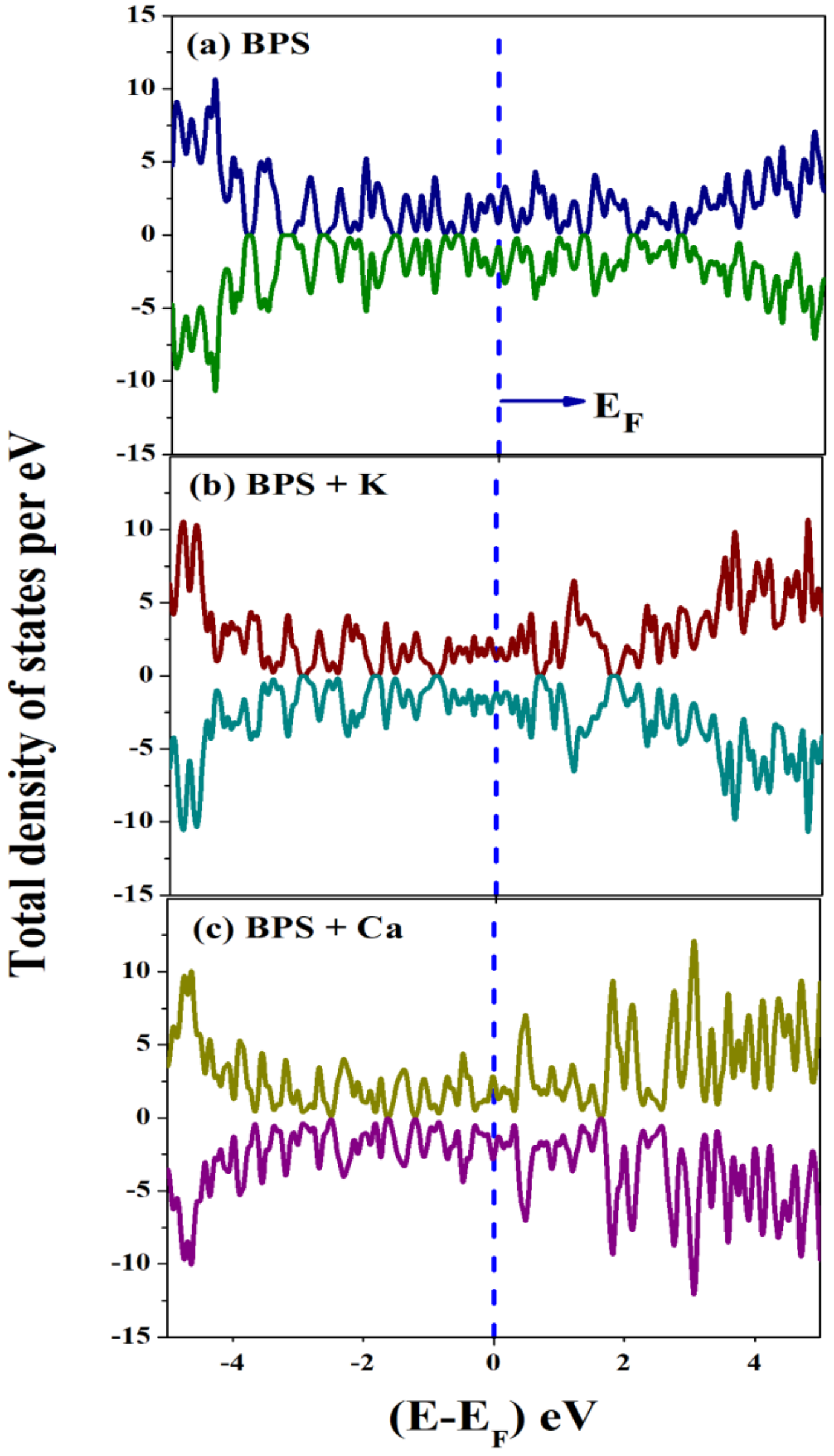

**Fig. 3. Total density of states of (a) Pristine BPS structure (b) K decorated BPS structure and (c) Ca decorated BPS structure. Fermi level is set at 0 eV. Upper and lower panel denote the total density of states of up and down spin, respectively.**

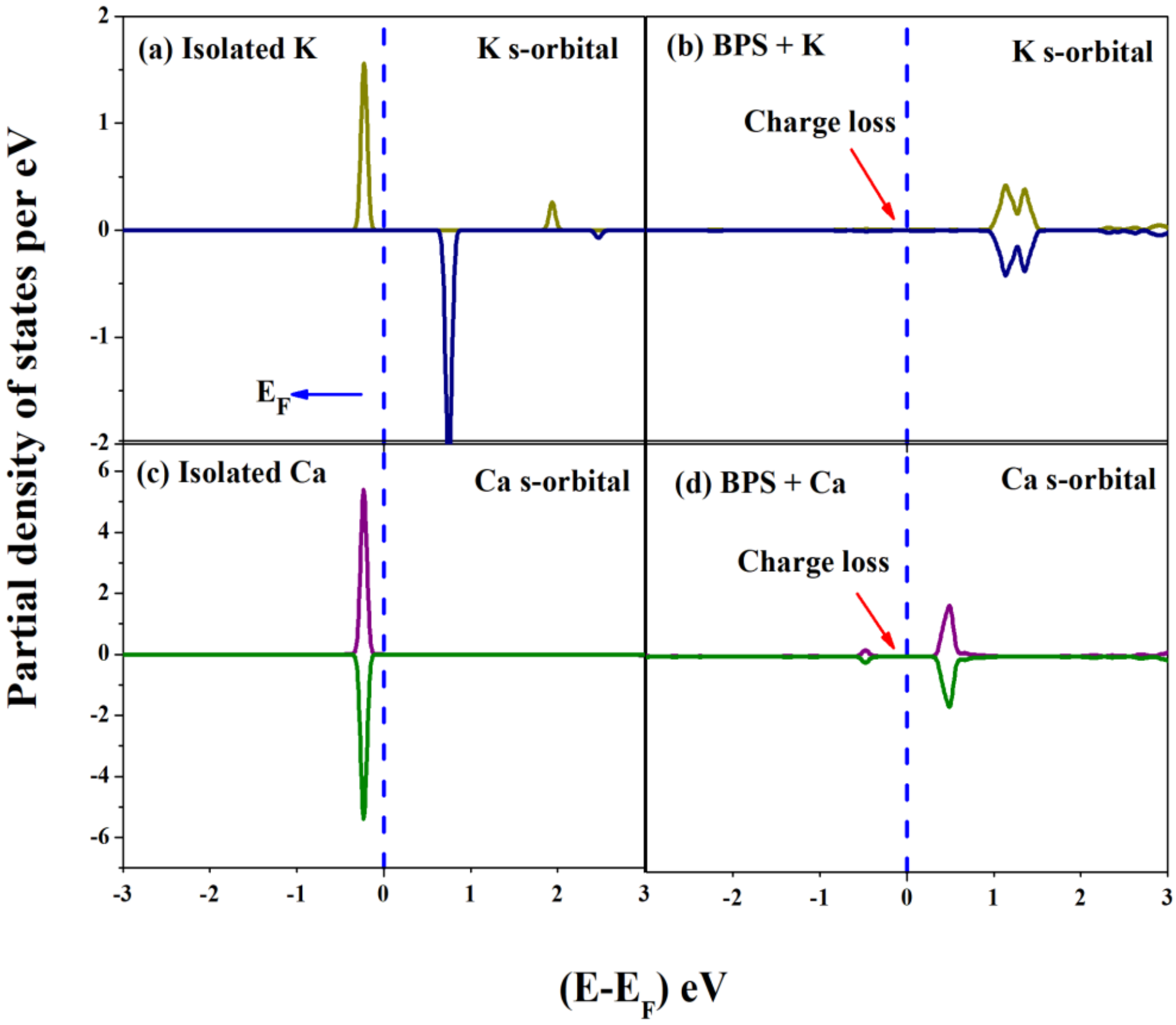

**Fig. 4. Partial density of states of (a) s-orbital of isolated K atom (b) K-s orbital of BPS + K structure (c) s-orbital of isolated Ca atom (d) Ca-s orbital of BPS + Ca structure. Fermi level is set at 0 eV.**

## Bader charge analysis

DOS and PDOS analysis explain the charge transfer process in a qualitative manner. For quantitative understanding, to get an approximated value of charge, which has been transferred from the metal atoms to the BPS structure, we have carried out the Bader charge analysis[79] calculations. From the Bader charge analysis, we have found that a total amount of 0.90e and 1.67e charges have been transferred from K and Ca atoms to the BPS structure. Ca atom transfers more charge to BPS due to its lower ionization potential as compared to K atom. Since the valence shell electronic configuration of the K and Ca atom is $4s^1$ and $4s^2$, respectively, more charge has been transferred from the valence shell of the Ca atom to the BPS, compared to the K atom. As a result, the Ca atom is bonded with more binding energy to BPS than the K atom. We have found that a significant amount of charge has been transferred from the metal atoms to the BPS structure, due to which metal atoms are bonded strongly with the BPS structure.

**Spatial charge density plot**

We present the spatial charge density difference plots to depict the charge transfer between the metal atoms and the BPS structure. The top and side views of spatial charge density difference between ρ (BPS + K) – ρ (BPS) for iso-surface value 0.0023e are presented in **Fig. 5 (a & b).** The spatial charge density difference plots for ρ (BPS + Ca) – ρ (BPS) are presented in **Fig. 5 (c & d).** The iso-surface value for **Fig. 5 (c &d)** is 0.0052e.

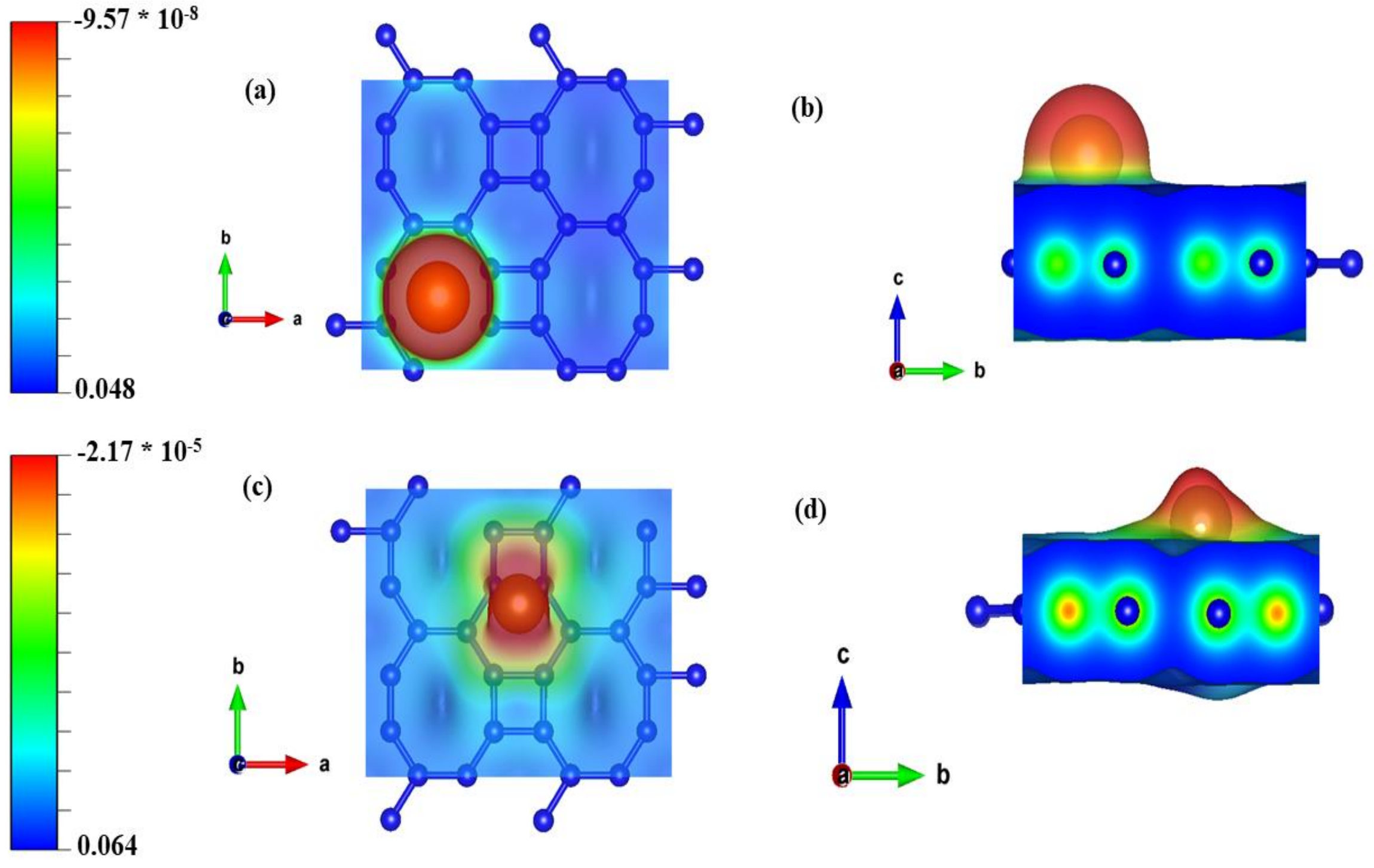

**Fig. 5. Electronic charge density difference plots for (a) Top view of ρ(BPS + K) – ρ(BPS) system for isosurface value 0.0023e. (b) side view of ρ(BPS + K) – ρ(BPS) system (c) Top view of ρ(BPS + Ca) – ρ(BPS) system for isosurface value 0.0052e. (d) side view of ρ(BPS + Ca) – ρ(BPS) system. The plots are in B-G-R color pattern, red and green colors denote charge loss and charge gain regions respectively.**

These plots correspond to the B-G-R color pattern, in which the red color around the metal atoms denotes the charge loss while the green color denotes the charge gain region. It is clear from **Fig. 5 (a & b)** that when the K atom is attached on the top of the center of the octagon of BPS, the most of the charge has been gained by the carbon atoms of the octagon, while since the Ca atom is attached on the top of the common face of tetragon and hexagon, most of the charge has been gained by the carbon atoms of the nearest tetragon and hexagon as displayed in **Fig. 5 (c & d).** Therefore, some charge has been transferred from the K and Ca atoms to BPS, responsible for the strong binding of metal atoms to the BPS structure. The charge density plots are consistent with the partial density of states and Bader charge analysis.

### 3.2 Hydrogen adsorption on K and Ca decorated BPS structures

The hydrogen molecules are kept at almost 2 Å distance above the minimal energy sites of metal decorated BPS structures, and relaxation calculations are performed. We have corrected the generalized gradient approximation functional results by employing the van der Waals corrections of DFT-D2 type[84,85] to consider the effect of van der Waals interactions for accurate binding energy calculations. The metal atoms attached to the BPS structure are positively charged due to their charge transfer to the BPS structure. When hydrogen molecules are kept in the vicinity of these metal cations, they get polarized [54,58]. Therefore, the induced electric field between the metal atoms (K and Ca) and $H_2$ molecules is responsible for the hydrogen binding to metal decorated BPS structure. In addition to the electrostatic interactions, orbital hybridization also plays a significant role in the hydrogen adsorption. Ataca et al.[39] have investigated hydrogen adsorption in alkali-earth metals (Be, Mg, Ca) decorated graphene system and found that orbital hybridization between the 3d orbitals of Ca atom and $\pi^*$ antibonding orbitals of graphene are responsible for the strong binding of Ca atom on graphene. Lee et al.[54] have reported that the hybridization of the Ca empty 3d orbitals filled $\sigma$ orbitals of $H_2$ are responsible for hydrogen adsorption. Chen et al.[61] have also shown that the Kubas interaction between the empty 3d orbitals of Ca atom and filled molecular orbital of $H_2$ plays a crucial role in hydrogen adsorption. To investigate whether Kubas interaction[51], in which back charge donation from filled σ orbitals of $H_2$ to vacant 3d orbitals of metal atoms (K and Ca) take place, is also responsible for the binding of hydrogen, we have presented the partial density of states of the 3d orbitals of K and Ca atoms before and after the addition of $1^{st}$ $H_2$ molecule in **Fig. 6.**

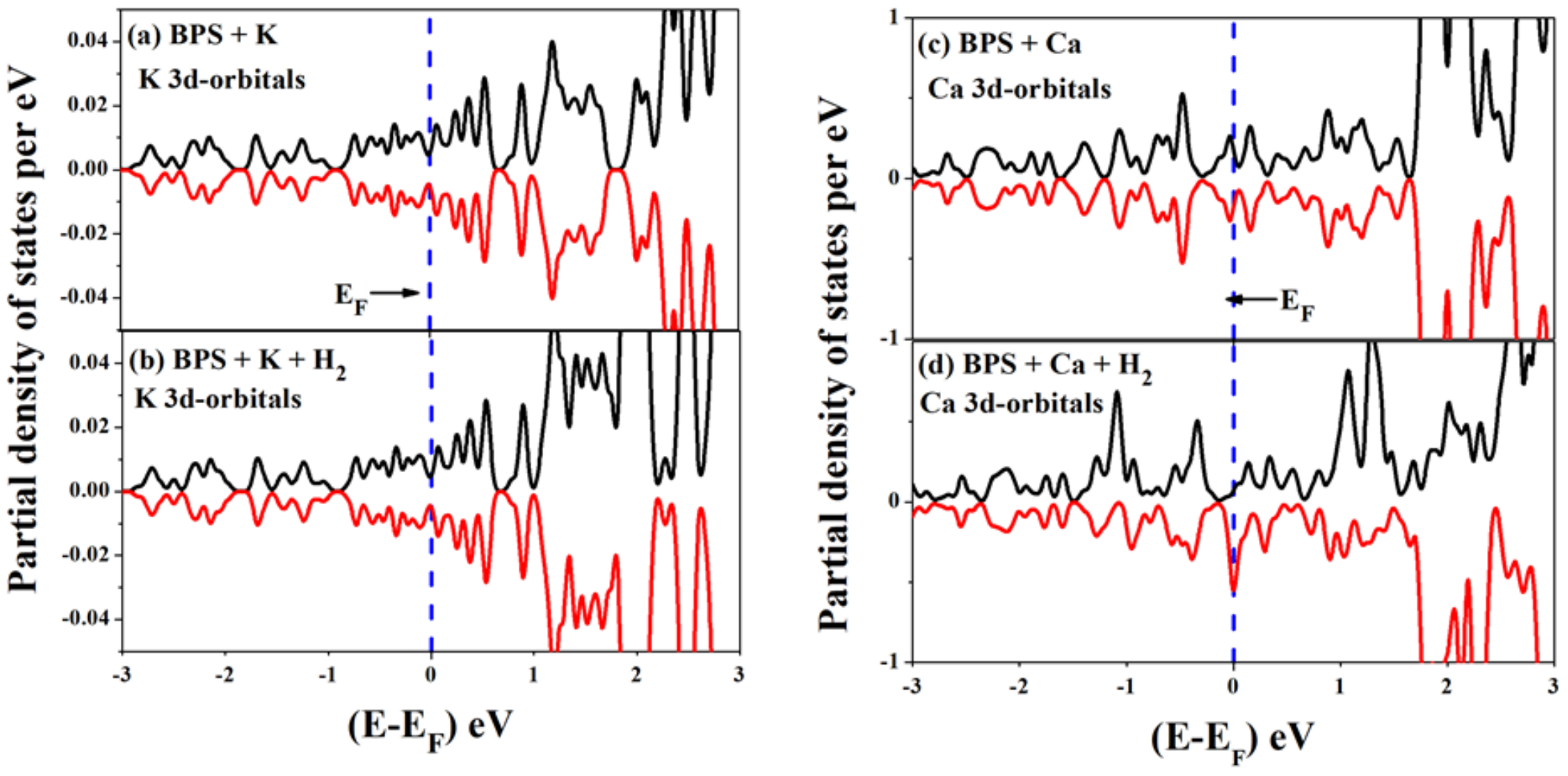

**Fig. 6. Partial density of states of (a) K 3d-orbitals for BPS + K system (b) K 3d-orbitals for BPS + K +$H_2$ system (c) Ca 3d-orbitals for BPS + Ca system (d) Ca 3d-orbitals for BPS + Ca + $H_2$ system. Fermi level is set at 0 eV.**

We have found that the partial density of the states of 3d orbitals of K atom is similar near Fermi energy level before and after the addition of $1^{st}$ $H_2$ molecule in the BPS + K structure, as shown in **Fig. 6 (a & b)**. Therefore, orbital hybridization of K 3d-orbitals and σ orbitals of $H_2$ is not involved in the binding of $H_2$ molecule with K atom. However, from **Fig. 6 (c & d),** it is clear that the nature of partial density of states of 3d orbitals of Ca atom changes near Fermi energy level, when $H_2$ molecule is attached to BPS + Ca structure. This implies that Kubas interaction is also responsible for the binding of $H_2$ molecules with Ca atom of BPS + Ca structure, in addition to electrostatic and van der Waals interactions.

Initially, we kept the 1st $H_2$ at almost 2 Å distance from the K and Ca atoms. The binding energies of the 1st $H_2$ are -0.20 eV and -0.26 eV for K and Ca decorated BPS structures, respectively. The negative binding energy represents exothermic reactions, which are related to the stability of the system. The adsorption energy of the nth $H_2$ molecule was determined by using the following equation:

$$\boldsymbol{E_n(H_2) = E\ (BPS + M + n\ H_2) - E(BPS + M + (n-1)\ H_2) - E\ (H_2)} \qquad (2)$$

Where ***E (BPS + M + n $H_2$) and E (BPS + M + (n-1) $H_2$)*** are the total energy of metal decorated BPS structure with n and (n-1) $H_2$ molecules, respectively, and ***E($H_2$)*** is the energy of the isolated hydrogen molecule.

After determining the relaxed structure of BPS + M + $H_2$, we have added more hydrogen molecules successively and found that K and Ca decorated BPS structure can adsorb 5 $H_2$ with the adsorption energy in the range for reversible hydrogen storage as specified by the DOE-US[8,9]. The adsorption energy of the 6th $H_2$ molecule is lesser than 0.2 eV for both K and Ca decorated BPS structures, which is not suitable for the reversible adsorption of hydrogen. The average binding energies of the adsorbed five hydrogen molecules for K and Ca decorated BPS structures are -0.24 eV/$H_2$ and -0.33 eV/$H_2$, respectively, which are suitable for reversible hydrogen storage energy range (-0.20 eV to -0.40 eV). Denis et al.[78] have reported average binding energy of -0.20 eV/$H_2$ for Li decorated biphenylene sheet using van der Waals incorporated density functional theory. Lee et al.[58] have reported -0.20 eV/$H_2$ average binding energy of hydrogen molecules attached on Ca decorated zigzag graphene nanoribbons (ZGNR) structure.

We have found that the H-H bond distance for the 1st hydrogen molecule is 0.75 Å for BPS + K + 1 $H_2$ structure, which is very close to the H-H bond distance of isolated $H_2$ of 0.74 Å. The H-H bond distances are around ~0.75 Å for all the adsorbed hydrogen molecules on the BPS + K structure. The H-H bond distance is 0.76 Å for the 1st hydrogen molecule attached to the Ca atom of the BPS + Ca structure, and the H-H bond distances are found to be in a range of 0.76 Å – 0.77 Å for all the adsorbed hydrogen molecules on BPS + Ca structure. Slightly more elongation in the H-H bond lengths takes place when hydrogen molecules are attached to the Ca atom compared to the K atom because $H_2$ molecules are bounded with electrostatic and Kubas interaction to the Ca atom while the Kubas interaction is absent in the case of K decorated BPS structure. The optimized structures of BPS + K + (n)$H_2$ and BPS + Ca + (n)$H_2$ compositions for (n = 1 to 5) are shown in **Fig. (7)** and **Fig. (8),** respectively.

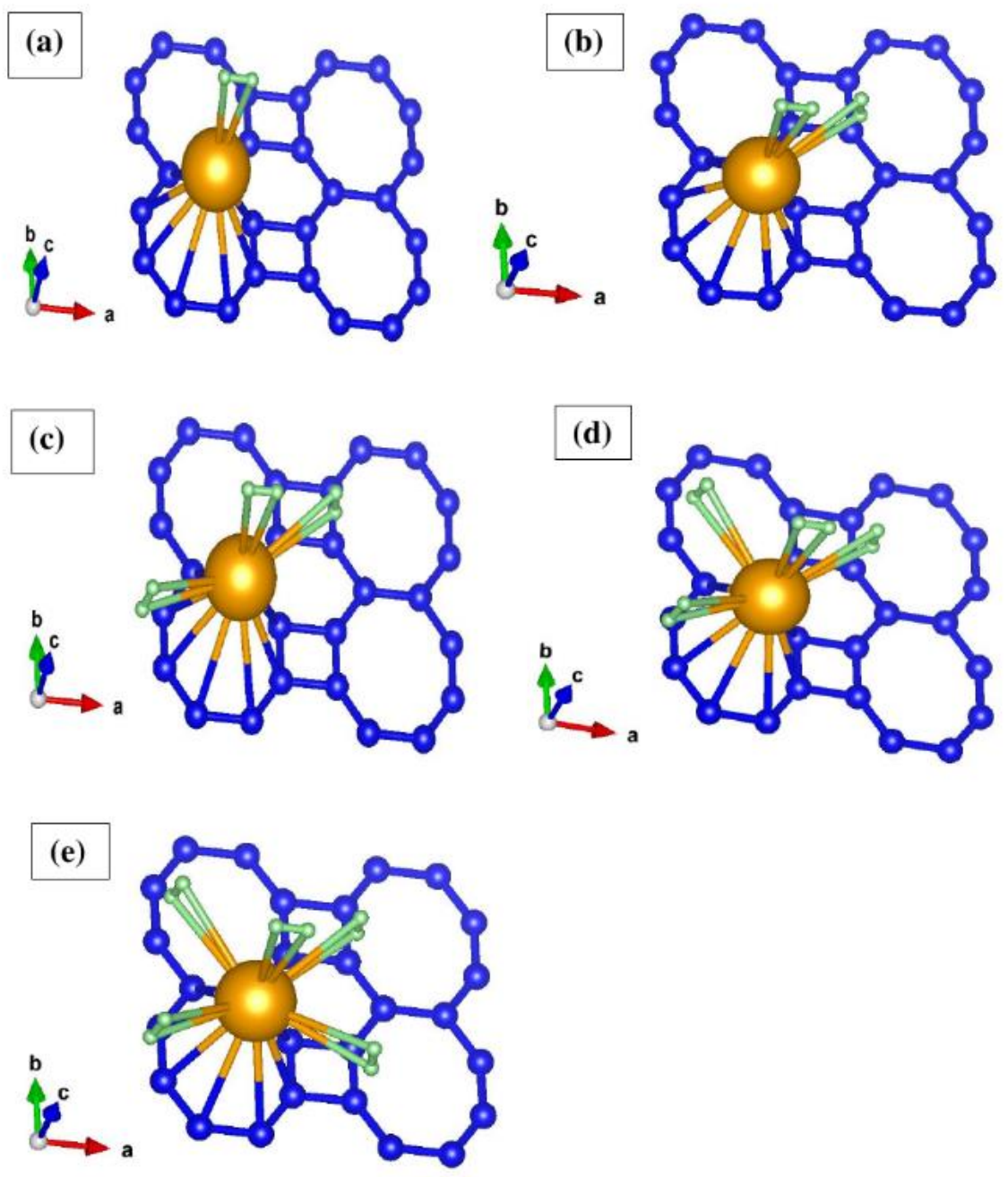

**Fig. 7. Optimized structures of (a) BPS + K + $H_2$ (b) BPS + K + 2$H_2$ (c) BPS + K + 3$H_2$ (d) BPS + K + 4$H_2$ (e) BPS + K + 5$H_2$ compositions. Here blue, orange and green colors denote the carbon, potassium, and hydrogen atoms, respectively.**

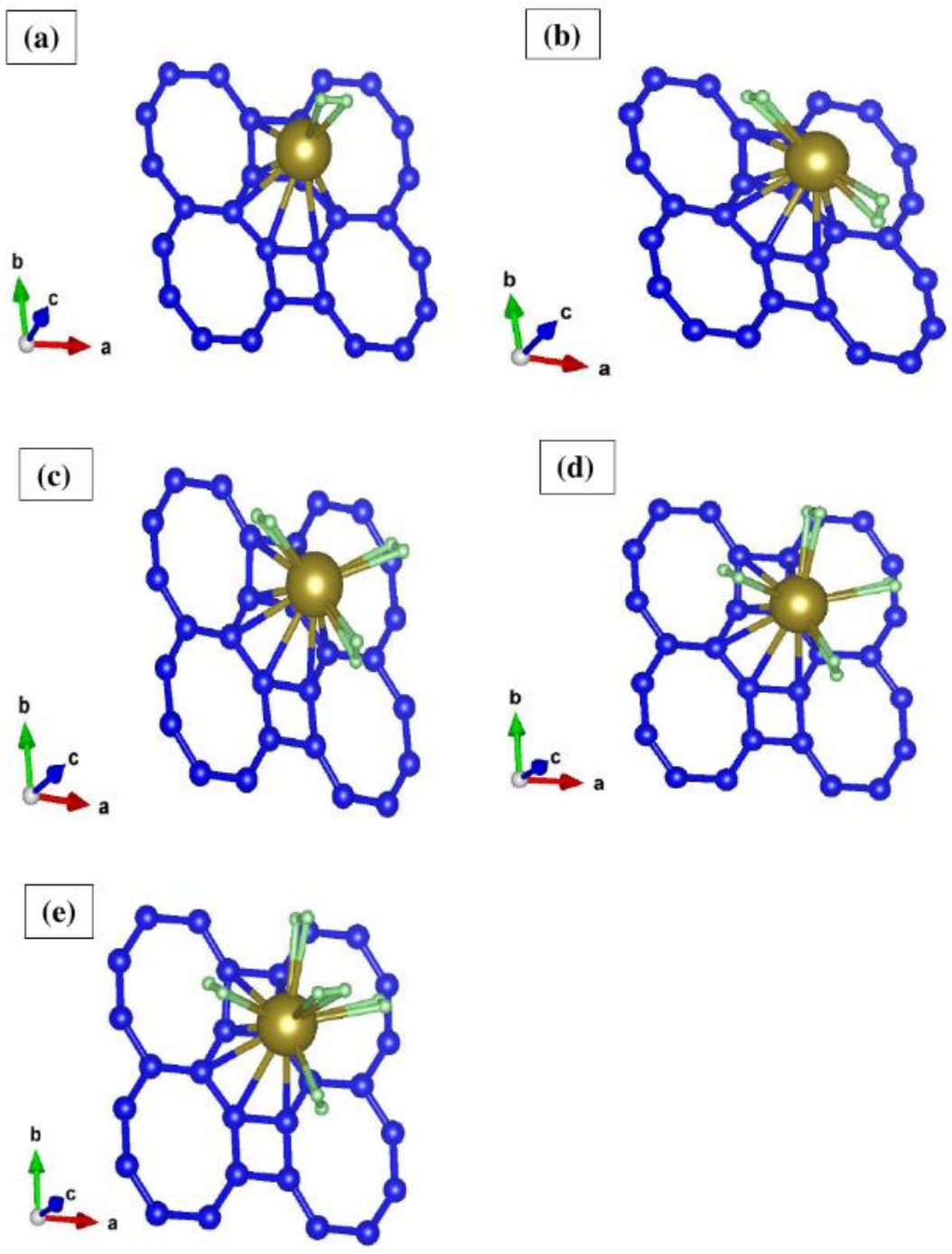

**Fig. 8. Optimized structures of (a) BPS + Ca + $H_2$ (b) BPS + Ca + $2H_2$ (c) BPS + Ca + $3H_2$ (d) BPS + Ca + $4H_2$ (e) BPS + Ca + $5H_2$ compositions. Here blue, golden and green colors denote the carbon, calcium and hydrogen atoms respectively.**

The adsorption energies of the $H_2$ are provided in **Table 2.**

**Table 2. Adsorption energy for $H_2$ molecules attached on K and Ca decorated BPS structures using GGA + DFT-D2 method.**

| **Compositions**<br>BPS + M + (n)$H_2$<br>M = K, Ca and n= 1 to 5 | **Adsorption energy (eV)**<br>**(GGA + DFT-D2)**<br>**BPS + K** | **Adsorption energy (eV)**<br>**(GGA + DFT-D2)**<br>**BPS + Ca** |
|---|---|---|
| BPS + M + $H_2$ | -0.20 | -0.26 |
| BPS + M + $2H_2$ | -0.28 | -0.48 |
| BPS + M + $3H_2$ | -0.26 | -0.35 |
| BPS + M + $4H_2$ | -0.22 | -0.34 |
| BPS + M + $5H_2$ | -0.22 | -0.24 |
| Average adsorption Energy per $H_2$ | -0.24 | -0.33 |
| Average desorption Temperature | 310 K | 425 K |
| Gravimetric wt % | 11.90 | 11.63 |

**3.3 Estimation of desorption temperature and gravimetric weight percentage (wt %) of hydrogen**

The adsorbed $H_2$ molecules should get desorb from the metal decorated host structure at suitable temperatures for practical use. The $H_2$ molecules should remain attached to the host structure at ambient conditions, and the releasing temperature of $H_2$ molecules should not be significantly higher than room temperature otherwise, one has to supply additional energy to use the adsorbed hydrogen molecules for practical applications. We have calculated the average desorption temperature of the $H_2$ by employing the Van't Hoff equation[12]:

$$\boldsymbol{T_d} = \left(\frac{\boldsymbol{E_b}}{\boldsymbol{k_B}}\right)\left(\frac{\Delta S}{\boldsymbol{R}} - \boldsymbol{ln\ P}\right)^{-1} \qquad (\mathbf{3})$$

Here, $\boldsymbol{T_d}$ is the average desorption temperature. $\boldsymbol{E_b}$, $\boldsymbol{k_B}$, $\boldsymbol{\Delta s}$, $\boldsymbol{R}$, and $\boldsymbol{P}$ are the average binding energy of the adsorbed $H_2$, Boltzmann constant, entropy change for hydrogen for gas to liquid phase conversion[90], gas constant, and atmospheric pressure, respectively. We have found that the average desorption temperature ($T_d$) values are 310 K and 425 K for K and Ca decorated BPS structures, respectively. The estimated values of the $T_d$ are very much suitable for reversible hydrogen storage applications[40,91,92].

For the gravimetric weight percentage calculations, the clustering issue of the metal atoms should be taken care of. Although the K atom binds to the BPS with almost the same binding energy as on H and O sites (1.12 eV and 1.14 eV), we have placed the K atoms only above O sites of the BPS structure for the weight percentage calculations. We can put 4 K atoms above the center of the 4 octagons of BPS and 4 K atoms on the reverse side of the 4 octagons, as shown in **Fig. 9 (a & b)**.

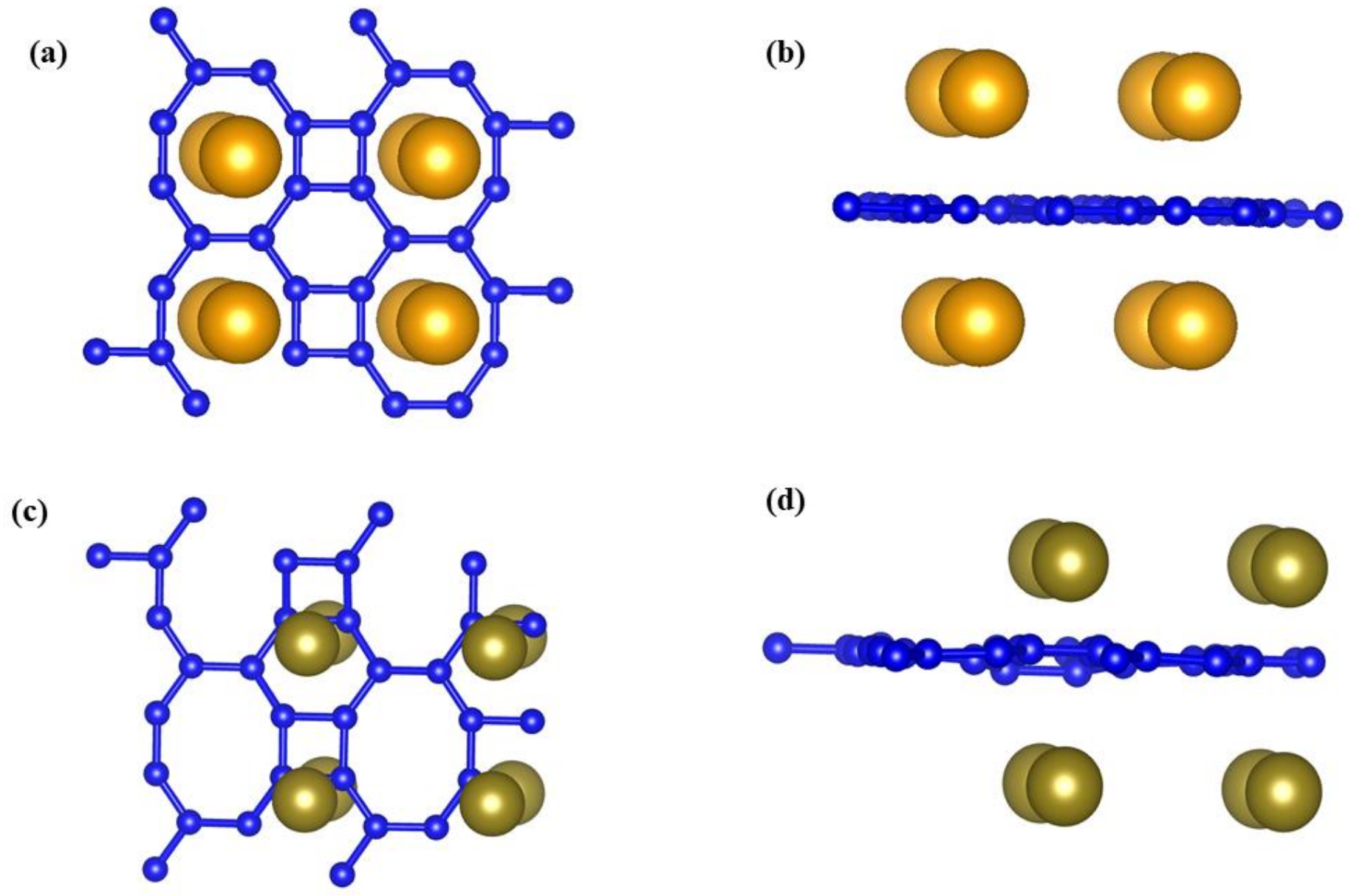

**Fig. 9. Metal loading pattern for hydrogen weight percentage calculations (a) Top view of BPS + 8 K atoms (b) Side view of BPS + 8 K atoms (c) Top view of BPS + 8 Ca atoms (d) Side view of BPS + 8 Ca atoms. Wt % of hydrogen for K and Ca decorated BPS structures are 11.90 % and 11.63 % respectively.**

Therefore, the 2×2×1 supercell of BPS structure can adsorb 8 K atoms, and each K can adsorb up to 5 hydrogen molecules. The gravimetric weight percentage for the K decorated BPS structure is 11.90 %, far greater than the DOE-US requirements.

Similarly, 2×2×1 supercell of BPS can adsorb 8 Ca atoms, with 4 above the common side of the tetragon and hexagon (maximum binding energy site for Ca atom) of the BPS structure, and 4 on the reverse side of the BPS structure as shown in **Fig. 9 (c & d).** We have also determined the energy barrier for the diffusion of Ca atom and performed the molecular dynamics simulations to explain the absence of clustering in the Ca decorated BPS system for this metal loading pattern as we know that the binding energy of Ca atom on BPS is in between to the cohesive energy of cluster and bulk Ca. Since each Ca atom can adsorb up to 5 $H_2$, the gravimetric weight percentage for the Ca decorated BPS structure is 11.63 %. We have found that the average binding energy of $H_2$, hydrogen release temperature, and gravimetric wt % of $H_2$ for K and Ca decorated BPS systems lie in the suitable range for the reversible hydrogen storage system for practical applications. We have compared various hydrogen storage parameters of metal decorated BPS structure with some of the previous studies on different carbon nanostructures in **Table 3.**

**Table 3. Hydrogen storage parameters comparison for various carbon nanostructures.**

| **Metal doped system** | **Total no. of adsorbed Hydrogen molecules** | **Average adsorption energy per $H_2$ (eV)** | **Average desorption temperature (K)** | **Gravimetric wt % of $H_2$ (%)** |
|---|---|---|---|---|
| Graphene + Ca[39] | 5 | - | - | 8.4 |
| Graphyne + Li[43] | 4 | -0.27 | - | 18.6 |
| Graphdiyne + Li[93]<br><br>Graphdiyne + Na | 5<br><br>5 | -0.28<br><br>-0.25 | -<br><br>- | 8.81<br><br>7.73 |
| $B_{80}$ + Ca[8] | 12 | -0.12 - 0.40 | | 8.2 |
| HGY + Li[47] | 4 | -0.22 | 353 | 12.8 |
| BPS + Li[78] | 7 | -0.20 | - | 7.4 |
| ***BPS + K***<br><br>***BPS + Ca***<br><br>***(Present Work)*** | **5**<br><br>**5** | **-0.24**<br><br>**-0.33** | **310**<br><br>**425** | **11.90**<br><br>**11.63** |
| **Experimental** | | | | |
| MWCNTs + Pd[94] | **-** | **-** | **-** | 6.0 |
| Graphene + Ni + Al[95] | - | - | - | 5.7 |

### 3.4 Practical viability of the system

#### Diffusion energy barrier calculations

The binding energy of the Ca to the BPS structure (1.41 eV) is more compared to the cohesive energy of cluster Ca (1.30 eV), so the possibilities of clustering in the Ca decorated BPS system is very small. But since the binding energy of Ca is less than the cohesive energy of bulk Ca, we have calculated the diffusion energy barrier for the displacement of Ca atom from one stable adsorption site to the nearest stable adsorption site by using climbing-image nudged elastic band (CI-NEB) method. We have found that there exists an energy barrier of 2.52 eV for the displacement of Ca atoms, as shown in **Fig. 10**.

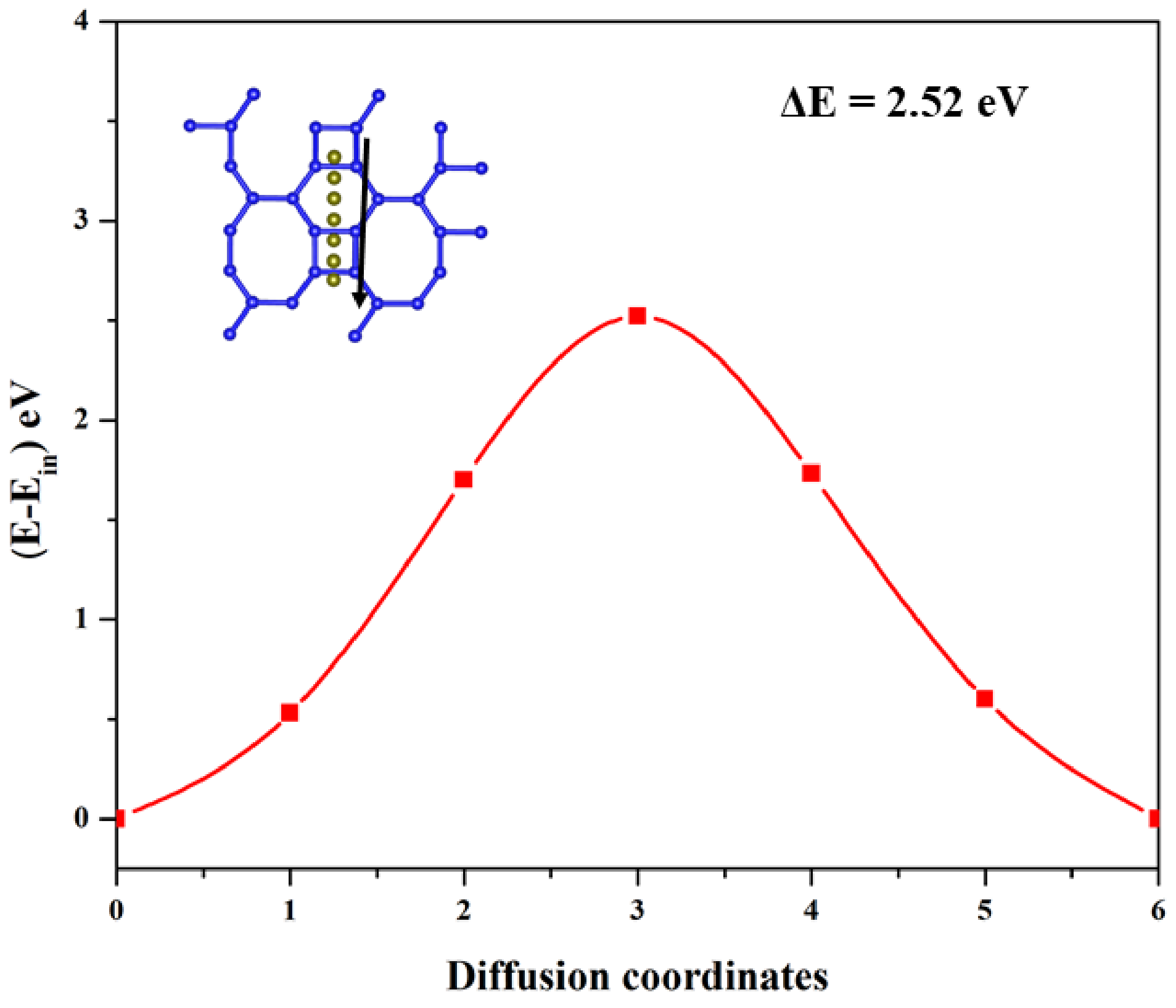

**Fig. 10. Diffusion energy barrier plot for the movement of the Ca-atom using CI-NEB method. Energy difference of current step energy and initial energy is plotted with respect to the small displacements of Ca atom.**

The energy barrier for the metal atoms should be more than the thermal energy of metal to restrict the movement of metal[47]. The thermal energy for Ca atom at desorption temperature 425 K is 0.055 eV, calculated using the formula:

$$E = \frac{3}{2} k_B T \qquad (4)$$

The diffusion energy barrier for the Ca atom for its movement from one site to the nearest stable site is 2.52 eV, which is much more than the highest thermal energy of the Ca atom 0.055 eV. Therefore, Ca atoms clustering should not occur in the Ca decorated BPS structures.

**The solidity of the metal decorated BPS structures at the desorption temperature**

For a practically viable hydrogen storage system, the metal atoms should remain attached to the BPS at high temperatures. We have carried out AIMD simulations for BPS + K and BPS + Ca structures and confirmed the solidity of the metal decorated BPS structures at their desorption temperatures. Initially, we have slowly increased the temperature of BPS + K and BPS + Ca systems up to their desorption temperatures of 310 K and 425 K, respectively, by putting the structures in the micro canonical ensemble for 5 ps time duration. The temperatures were increased in the time step of 1 fs. Then we have kept these systems in the canonical ensemble for another 5 ps time duration at their desorption temperatures.

The DFT optimized structure of BPS + K and the final molecular dynamics snapshot of BPS + K are shown in **Fig. 11 (a)** and **Fig. 11 (b),** respectively. Similarly, the optimized structure and the final molecular dynamics snapshot of the BPS + Ca system are presented in **Fig. 11 (c & d).** At desorption temperatures, K and Ca atoms move slightly from their equilibrium positions but remain attached to the BPS structure. The changes in metal to carbon distances are negligible. To further investigate the clustering issue in Ca decorated BPS structure, we have performed the molecular dynamics simulations for BPS + 2 Ca system. Initially, we performed the relaxation calculations of the BPS + 2 Ca system. The relaxed geometry of BPS + 2 Ca is shown in **Fig. 12 (a),** and the Ca-Ca bond distance was found to be 3.7 Å in the system. After determining the relaxed geometry of BPS + 2 Ca system, we performed the molecular dynamics simulations. The temperature of the BPS + 2 Ca system was increased up to 425 K in 5 ps time duration, and then the system was kept in a canonical ensemble for the next 5 ps time duration.

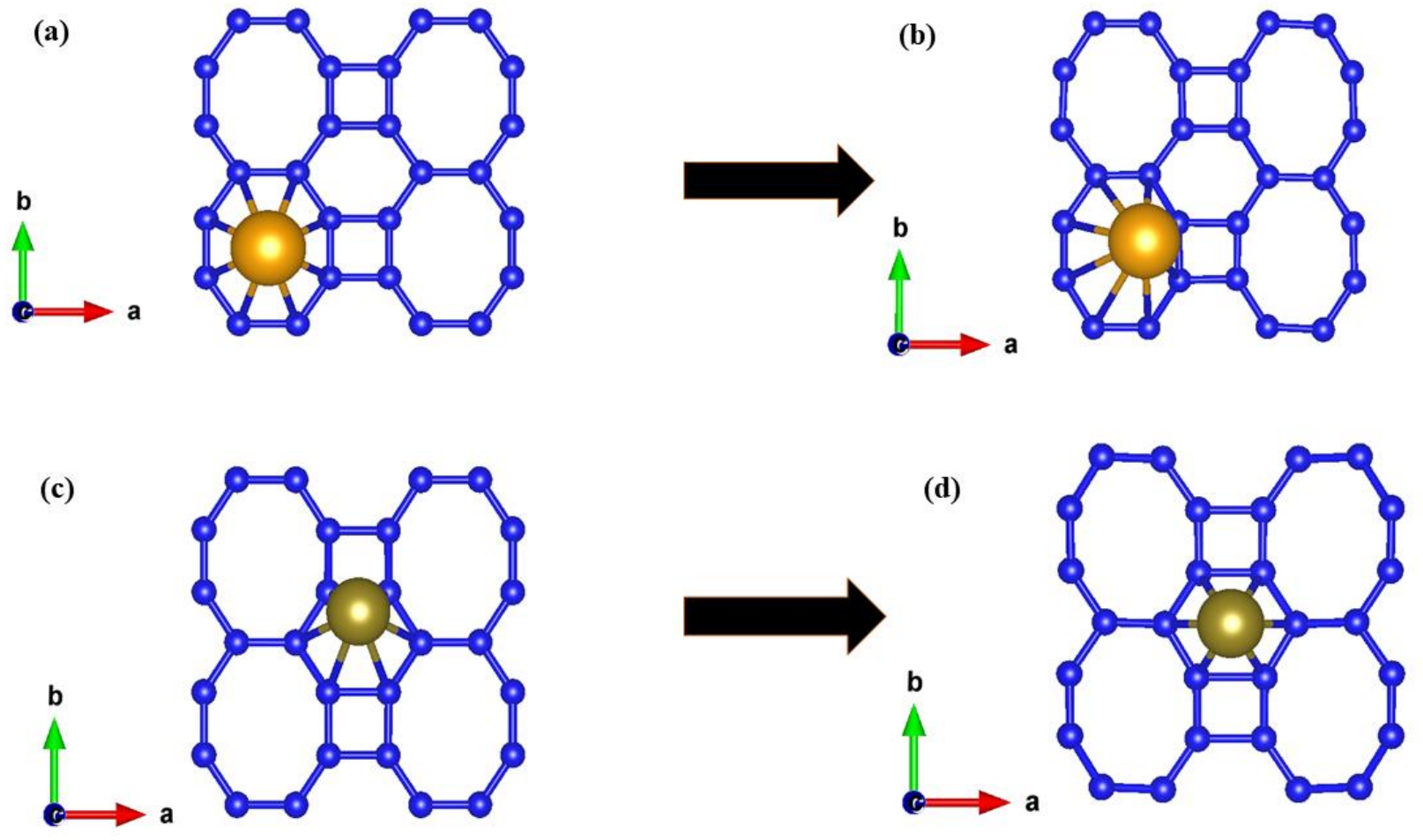

**Fig. 11 (a) Optimized structure of BPS + K system (b) MD snapshot of BPS + K system, after putting the system in canonical ensemble at 310 K for 5 ps time duration (c) Optimized structure of BPS + Ca system (d) MD snapshot of BPS + Ca system after putting the system in canonical ensemble at 425 K for 5 ps time duration. The changes in metal-carbon bond lengths are negligible.**

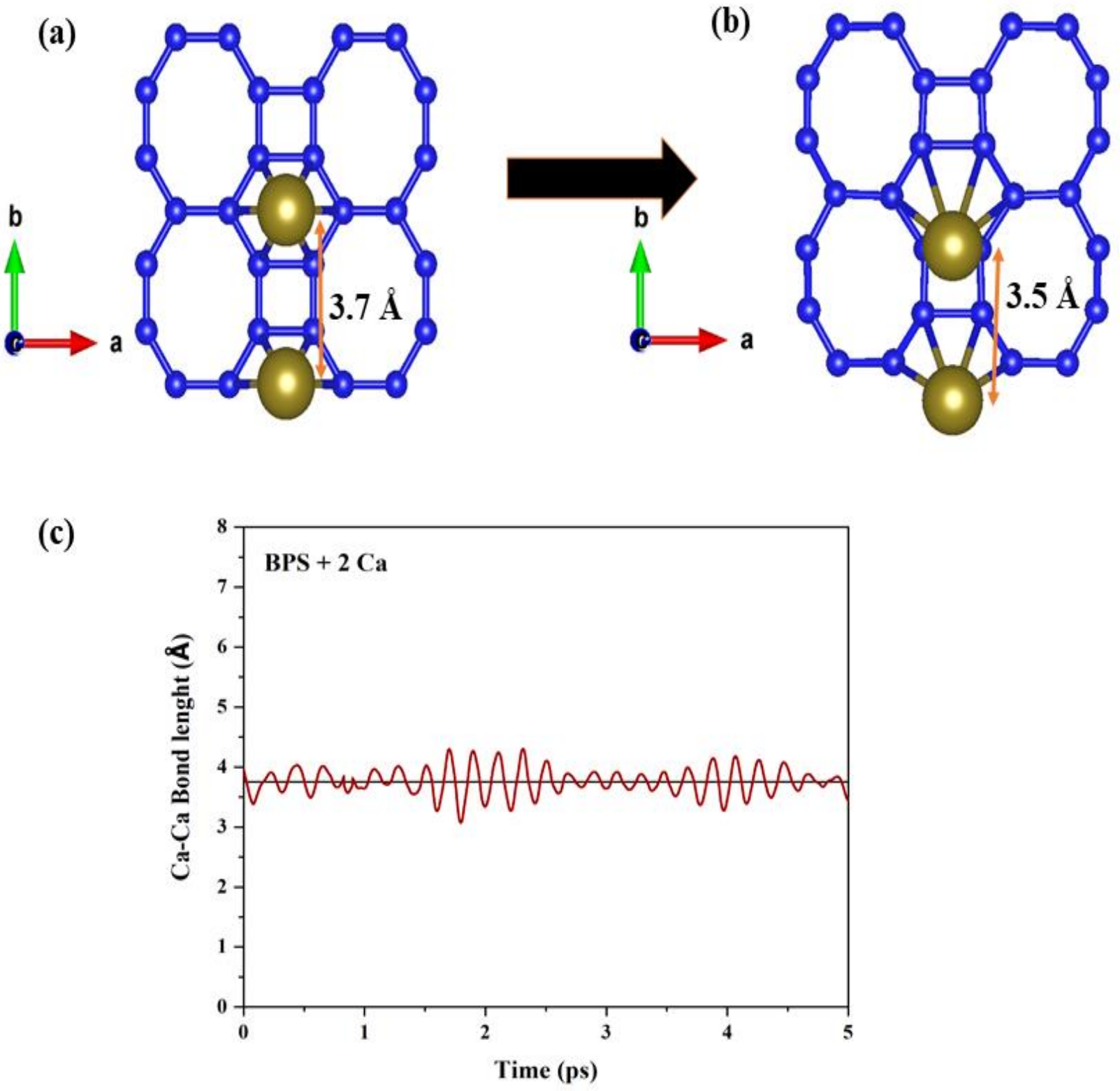

**Fig. 12. (a) Optimized structure of BPS + 2 Ca system (b) MD snapshot of BPS + 2 Ca system, after putting the system in canonical ensemble for 5 ps at 425 K (c) The Ca-Ca bond length fluctuations for BPS + 2 Ca system with the time duration of molecular dynamics simulations.**

The final molecular dynamics snapshot of the BPS + 2 Ca system is presented in **Fig. 12 (b).** We have found that both the Ca atoms move slightly downward at the desorption temperature, and the Ca-Ca bond distance changes to 3.5 Å. The change in Ca-Ca bond distance at the desorption temperature is small (0.2 Å). We have also plotted the Ca-Ca bond distance with respect to the time of molecular dynamics simulations in **Fig. 12 (c).** The maximum Ca-Ca bond length fluctuations in the BPS + 2 Ca system are around ~0.5 Å, indicating that the Ca-Ca clustering should not occur. Since the metal atoms (K and Ca) remain attached to the BPS structure even at the desorption temperature, and the changes in the Ca-Ca bond length in BPS + 2 Ca system are small; hence we believe that K and Ca decorated BPS are practically viable hydrogen storage systems.

## 4. Conclusions

We have studied the hydrogen adsorption, and desorption behavior of AMs (Na, K) and AEMs (Be, Mg, Ca) decorated BPS structures. We have found that K and Ca atoms are strongly bonded to the BPS structure due to the significant amount of charge transfer from metal atoms to BPS. We have found that 5 $H_2$ can be attached on K and Ca decorated BPS with appropriate binding energy and desorption temperature for reversible hydrogen storage. We report 11.90 % and 11.63 % of hydrogen uptake for K and Ca decorated BPS, respectively, which is much higher than the DOE criterion of 6.5 %. We have investigated the clustering issue for Ca atom as the binding energy of Ca on BPS is lesser than the cohesive energy of bulk Ca, and found that the sufficient amount of energy barrier will restrict the clustering. The AIMD simulations explain the integrity of K and Ca decorated BPS structures at high temperatures and the absence of metal-metal clustering. The average binding energy ($E_b$) and releasing temperature ($T_d$) of K and Ca decorated BPS is suitable for reversible hydrogen storage, the weight percentage of hydrogen is significantly higher than the DOE-US guidelines, and metal decorated BPS structures are stable at desorption temperatures. Therefore, we strongly believe that the K and Ca decorated BPS structures are practically suitable, ultrahigh-capacity, reversible hydrogen storage candidates, and our results will motivate the experimentalist to explore the hydrogen storage properties of K and Ca decorated BPS structures.

## Acknowledgment

VM would like to acknowledge DST-INSPIRE for providing the fellowship and SpaceTime-2 supercomputing facility at IIT Bombay for the computing time. BC would like to thank Dr. T. Shakuntala and Dr. Nandini Garg for support and encouragement. BC also acknowledge support from Dr. S.M. Yusuf and Dr. A. K Mohanty.